\documentstyle{l-aa}    

\begin{document}    
    
\thesaurus{02.01.1, 02.19.1, 09.03.2, 09.19.2}    
    
\title{Supernova remnants in molecular clouds: on cosmic ray electron 
spectra}    
    
\author{M. Ostrowski}    
    
\institute{Obserwatorium Astronomiczne, Uniwersytet Jagiello\'nski,    
ul.Orla 171, 30-244 Krak\'ow, Poland}    
    
\offprints{E-mail: mio\@@oa.uj.edu.pl)}    
    
\date{Received ...; accepted .. ; }    
    
\maketitle     
    
\begin{abstract}    
The particle acceleration process at a shock    
wave, in the presence of the second-order Fermi acceleration in the    
turbulent medium near the shock, is discussed as an alternative    
explanation for the observed flat synchrotron spectra of supernova    
remnants (SNRs) in molecular clouds. We argue that    
medium Alfv\'en Mach number shocks considered by Chevalier (1999)    
for such SNRs can naturally lead to the observed spectral indices.    
    
    
\end{abstract}    
    
\section{Introduction}    
    
In a recent paper Chevalier (1999) discusses physical conditions in    
supernova remnants (SNRs) evolving inside molecular clouds. He considers    
a clumpy gas distribution in the clouds, where the dominant    
part of the cloud mass is contained in the compact dense clumps filling    
only 10\% or less of the cloud volume and the rest of the cloud is    
filled with tenuous gas with a number density $N \sim 10$ cm$^{-3}$. Thus a    
SNR exploding inside a cloud evolves mostly in such low density    
medium. Chevalier discusses a number of consequences of such a model and    
compares it to observations of three SNRs: W44, IC 433 and 3C391. In    
his discussion the observed, very flat, cosmic ray electron distributions    
responsible for the radio synchrotron spectra are interpreted as    
shock-compressed ambient distributions radiating downstream of the    
radiative shock.    
    
However, the existence of a very flat ambient electron distribution in 
a molecular cloud is a matter for debate. The uniform magnetic field 
structures observed in several clouds and efficient damping of short 
Alfv\'en waves in a partly neutral medium (Hartquist \& Morfill 1984) 
suggests a possibility of an efficient cosmic ray exchange between 
clouds and a steep-spectrum galactic cosmic ray population. In the 
present note we point out the possibility of an acceleration of the flat 
spectrum electrons at a SNR shock wave, if the second-order Fermi 
acceleration in the vicinity of the shock is taken into account. In the 
next section we summarize results of an approximate analytic theory for the 
particle spectral indices (Ostrowski \& Schlickeiser 1993, $\equiv$ 
`OS93'), with a few typing errors of the original paper corrected. Then, 
in section 3, we demonstrate that the physical conditions considered by 
Chevalier (199) for shock waves -- with the Alfv\'en velocity, $V_A$, 
non-negligible in comparison to the shock velocity -- allow for 
generation of the required flat particle distributions (c.f., also, 
Drury 1983, Hartquist \& Morfill 1983, Dr\"oge et al. 1987; Schlickeiser 
\& F\"urst 1989). In the final section (Section 4) we briefly discuss an 
application of the present model to actual conditions in astrophysical 
objects. 
 
\vfill 
    
\section{Derivation of the particle spectral index}    
    
OS93 derived a simplified kinetic equation for the particle distribution    
formed at the parallel shock wave due to action of the first-order    
acceleration at the shock and the second-order acceleration in the shock    
turbulent vicinity. With the phase-space distribution function at the    
shock, $f(p) \equiv f(p,x=0)$, an `integral' distribution is defined as 
 
\vfill 
    
$$F(p) = f(p) {\left[ {\kappa_1(p) \over U_1} + {\kappa_2(p) \over U_2 }    
\right]},\eqno(2.1)$$    
    
\vfill 
 
\noindent    
where $U_i$ and $\kappa_i$ are the plasma flow velocity in the shock     
frame and the spatial diffusion coefficient of cosmic ray particles,    
respectively ($i = 1$ upstream and $2$ downstream of the shock).    
Analogously, the momentum diffusion coefficient is indicated as $D_i$    
($i = 1$, $2$). For the simple case of $\kappa_i = const$ ($i =    
1,2$), and $D_1 = D_2 = D$ the approximate transport equation for the    
function (2.1) takes the closed form:    
 
\vfill \pagebreak 
    
$$-{1 \over p^2} {\partial \over \partial p} \left\{ p^2 D(p) {\partial    
\over \partial p} F(p) \right\} + {1 \over p^2} {\partial \over \partial    
p} \left\{ p^2 {\langle {\Delta p \over \Delta t} \rangle} F(p) \right\}    
$$    
    
$$ + {F(p) \over {\tau(p)}} = Q(p) \qquad , \eqno(2.2)$$    
    
\noindent    
where we have included the source term $Q(p)$. The mean acceleration    
speed due to the first order process at the shock $\langle {{\Delta p} \over    
{\Delta t}}\rangle$ and the mean escape time due to advection downstream of    
the shock $\tau (p)$ can be derived from the spatial diffusion equation    
(cf. OS93) as:    
    
$${\bigg< {{\Delta p} \over {\Delta t} } \bigg> = {{R-1}\over{3R}}    
{U_1^2\over{\kappa_1+R\kappa_2}} p } \qquad , \eqno(2.3)$$    
    
$${ \tau (p) = {{R(\kappa_1+R\kappa_2)}\over U_1^2} } \qquad ,    
\eqno(2.4)$$    
    
\noindent    
where the shock compression $R \equiv U_1/U_2$. 
In more general conditions with $\kappa_i = \kappa_i(p)$ the function    
$f(p)$ must be explicitly present in the kinetic equation:    
    
$$-{1 \over p^2} {\partial \over \partial p} \left\{ p^2 D_1(p)    
{\partial \over \partial p} \left[ f(p) {\kappa_1(p) \over U_1} \right]    
\right\} $$    
    
$$ -{\kappa_2(p) \over U_2} {1 \over p^2} {\partial \over \partial p}    
\left\{ p^2 D_2(p) {\partial \over \partial p} f(p) \right\} +  $$    
    
$${1 \over p_2} {\partial \over \partial p} \left\{ p^2 \langle {\Delta    
p \over \Delta t} \rangle f(p) {\kappa_{ef}(p) \over U_1} \right\} +    
{f(p) \kappa_{ef}(p) \over U_1 \tau(p)} = Q(p) \qquad . \eqno(2.5)$$    
    
\noindent    
Let us note in the above equation the asymmetry between the first two 
terms. If only the first-order acceleration takes place ( $D_1 = 0 = 
D_2$), the known solution $f(p) \propto p^\sigma$, with $\sigma = 
-3R/(R-1)$, is reproduced by Eq.~(2.5). Obtaining a solution for a more 
general situation may be a difficult task. However, if we consider 
momenta much above (or below) the injection momentum, and the power-law 
form for the diffusion coefficient $\kappa \propto p^\eta$, the solution 
is also a power-law. In such conditions, from Eq.~(2.5) one can derive 
the spectral index $\sigma$. The Skilling (1975) formula is used for the 
momentum diffusion coefficient, relating it to the spatial diffusion, 
$D(p) = V_A^2 p^2 / (9 \kappa(p))$ . For a given jump condition at the 
shock for the Alfv\`en velocity and a given value of $\eta$, the 
resulting spectral index depends on only two parameters, the shock 
compression ratio $R$ and the velocity ratio $V_{A,1}/U_1$. 
    
OS93 checked the validity range of the approximate equation (2.5) with 
the use of numerical simulations. For $\eta \ne 0$ the equation can be 
used for the range of parameters preserving $V_A << U$, while, for 
$\kappa = const$, it provides a quite reasonable description of the 
particle spectrum at all $V_A < U$. 
    
In order to find a power-law solution of Eq.~(2.5) let us assume the 
following forms for the diffusion coefficients ($i$ = $1$, $2$): 
${\kappa_i (p) = \kappa_{0,i} p^\eta} $~, ${D_i (p) = D_{0,i} 
p^{2-\eta}} $~, where the constants $\kappa_{0,i}$ and $D_{0,i}$ are related 
according to the formula $D_{0,i} = V_{A,i}^2/(9\kappa_{0,i})$. With 
these formulae and the power-law form for the distribution function, 
$f(p) = f_0 p^\sigma $~, equation (2.5) yields a quadratic equation 
for $\sigma$ : 
    
$${ a\sigma^2 + b\sigma + c = 0 } \qquad , \eqno(2.6)$$    
    
\noindent    
with coefficients:    
    
$${ a = -{R\over 9U_1^2} ( V_{A,1}^2 + R V_{A,2}^2 ) } \qquad ,    
\eqno(2.7)$$    
    
$$b = (3 + \eta) a + { R-1 \over 3} \qquad , \eqno(2.8)$$    
    
$$c = -\eta { RV_{A,1}^2 \over 3U_1^2 } + R  \qquad . \eqno(2.9)$$    
    
\noindent    
Equation (2.6) has two solutions valid for, respectively, particle 
momentum much below ($\sigma > 0$) and much above ($\sigma < 0$) the 
injection momentum. Only the later one is of interest in the present 
considerations. Let us also note that the energy spectral index often 
appearing in the literature is $\Gamma = \sigma - 2$ and the synchrotron 
spectral index $\alpha = (\sigma - 3)/2 = (\Gamma - 1)/2$~.

\section{Spectral indices in realistic SNRs}    
    
In the SNR model discussed by Chevalier (1999) the shock wave propagates 
inside an inhomogeneous cloud. In the cloud, the dense clumps occupy 
less than 10\% of the cloud volume and the shock evolution proceeds mostly 
in the tenuous inter-clump medium with a number density $N \sim 
10$ cm$^{-3}$ and a magnetic field $B \approx 2\cdot 10^{-5}$ G. Much 
stronger magnetic fields ($\sim 10^{-4}$ G) may occur further 
downstream in the radiative shock. The measured shock velocities are in 
the range $U_1 \approx 80$ -- $150$ km/s. With the notation $N_0 \equiv 
N/(10 {\rm cm}^{-3})$, $B_0 \equiv B/(2 \cdot 10^{-5} {\rm G})$ and 
$U_0 \equiv U_1/(100 {\rm km/s})$ we find the ratio 
    
$$ {V_{A,1} \over U_1} = 0.14 B_0 N_0^{-1/2} U_0^{-1} \qquad . \eqno(3.1)$$    
    
\noindent     
It can be of order $0.1$ for the parameters 
considered above. Thus, the second order acceleration process can 
substantially modify the energy spectrum of particles accelerated at the 
shock. In Fig-s~1,2 we present spectral indices derived from the formulae of 
the previous section for the `canonical' Kolmogorov value  
$\eta = 0.67$~. The shock compression ratio $R = 4.0$ for the high Mach 
number adiabatic shock is assumed. An efficient particle acceleration 
requires a high amplitude turbulence near the shock (cf. Ostrowski 1994) 
and oblique magnetic field configurations may occur there. Thus, the 
mean magnetic field downstream of the shock is $B_2 > B_1$ due to shock 
compression of both the uniform and the turbulent field components. 
Below we use an effective jump condition for the Alfv\'en velocity 
$V_{A,2} = V_{A,1}$ (in general $B_1 \le B_2 \le B_1 R$ and, 
respectively, $V_{A,1}/\sqrt{R} \le V_{A,2} \le V_{A,1} \sqrt{R}$). The 
results for $V_{A,1}/U_1 < 0.2$ are considered, where the analytic 
formulae provide an accurate approximation for the actual spectra. 
    
\begin{figure}                    
\vspace{7.0cm} 
\includegraphics{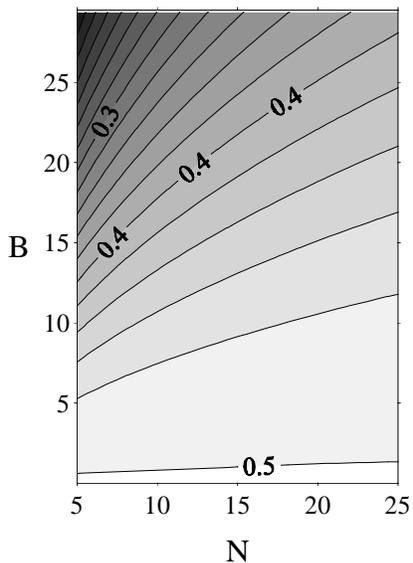}    
\caption[ ]{    
A map of the synchrotron spectral index $\alpha$ in co-ordinates of the    
magnetic field $B$ given in $\mu$G and the gas number density $N$ in    
cm$^{-3}$, both taken upstream of the shock.} \end{figure}    
    
For mean models for the SNRs discussed by Chevalier (1999)    
one has: \hfill\break    
a.) for W44: $N_0 = 0.5$, $U_0 = 1.5$, $\alpha = 0.33$    
($\Gamma = 1.72$)\hfill\break    
b.) for IC 443 (shell A): $N_0 = 1.5$, $U_0 = 0.8$, $\alpha = 0.36$    
($\Gamma = 1.66$) \hfill\break    
c.) for 3C391: $N_0 = 1.0$, $U_0 = 3.0$, $\alpha = 0.55$    
($\Gamma = 2.1$)\hfill\break    
With these parameters one can easily explain flat spectral indices 
for the cases (a) and (b) if reasonable values of the magnetic 
field are involved. Fig.~2 shows the spectral index 
$\alpha$ versus the magnetic field $B$ for the above listed choices of 
particle density $N = N_0$ and the shock velocity $U_1 = U_0$. The 
measured indices are indicated on the respective curves. The steep 
spectrum of 3C391 can not be exactly reproduced with the use of the simple 
model considered in this paper, as even neglecting the second order 
acceleration leads, for the assumed shock compression $R = 4.0$, to a 
somewhat flatter spectrum with $\alpha = 0.5$. However, the observed 
trend in the spectral index variation is well reproduced by the model. 
    
\begin{figure}                    
\vspace{5.7cm} 
\includegraphics{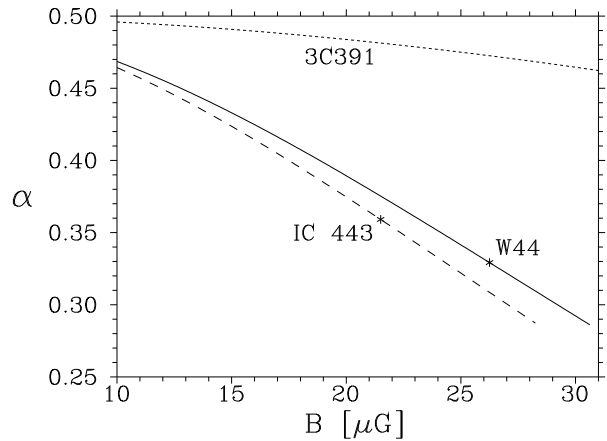} 
\caption[ ]{    
The synchrotron spectral index $\alpha$ versus the    
magnetic field $B$ for the respective values of $N$ and $U_1$. The    
curves for different objects are indicated with the respective SNR    
symbol. The observed spectral indices are indicated on the curves for IC 443    
and W44. } \end{figure}

\section{Discussion}    
    
A derivation of the particle spectral index at a shock front is    
presented for a situation involving the second-order acceleration    
process in the shock vicinity. We prove that even very weak shocks may    
produce very flat cosmic ray particle spectra in the presence of    
momentum diffusion. As a consequence, the dependence of the 
particle spectral index of the shock compression ratio can be    
weaker than that predicted for the case of pure first-order acceleration.    
    
One should note an important feature of the acceleration process in the    
presence of the second-order Fermi process acting near the shock.    
Because the same Alfv\'en waves scatter particle momentum in direction    
and in magnitude, there exists a strict link between the first- and the    
second-order acceleration processes. The particle spectrum is shaped    
depending on $V_{A,1}/U_1$, by the compression $R$, the 
momentum dependence of $\kappa$    
and, possibly, by the anisotropy of Alfv\'en waves determining the ratio    
of $\kappa$ to $D$. The last factor is not discussed here since under the    
present considerations we restrict ourselves to isotropic wave fields,    
but the role of anisotropy consists in decreasing the importance of momentum    
diffusion. Only through the above parameters the spectrum can be    
influenced by other physical characteristics of the shock and the medium    
in which it propagates. It is important that the Alfv\'en waves' amplitude,    
or the  magnitude of the spatial diffusion coefficient related to it, do    
not play any substantial role in determining the spectral inclination as    
both, first- and second-order, processes scale with it in the same way.    
One could argue that an energy transfer from waves to particles can cause    
quick damping of the waves and will leave us with the pure first-order Fermi    
acceleration at the shock. From the above discussion one can infer that    
such an objection is not valid for the isotropic wave field. Only the    
presence of high amplitude one-directional wave field enables the    
efficient first-order shock acceleration in the absence of momentum    
diffusion, but a detailed calculation is not possible without a detailed    
model for the upstream and downstream wave fields.    
    
In comparison to parallel shocks, the magnetic field inclined to the    
shock normal leads to a higher mean energy gain of particles interacting    
with the shock and a higher escape probability for downstream particles.    
Also, due to small cross-field diffusion the normal diffusive length    
scale near the shock decreases. As long as the shock velocity along the    
field is non-relativistic, the particle spectrum produced in the    
first-order process is not influenced by a field inclination. Because of    
smaller diffusive zones near the shock, the role of momentum diffusion    
may be of lesser importance as compared to the parallel case. This    
effect can be partly weakened by the presence of the magnetic field    
compression at the shock, which leads to higher Alfv\'en velocity    
downstream of the shock. Also the presence of high amplitude Alfv\'en    
waves (cf. Micha{\l}ek \& Ostrowski 1996) and an admixture of fast mode    
waves propagating obliquely with respect 
to the magnetic field (Schlickeiser \& Miller    
1998, Micha{\l}ek et al. 1998) may lead to a more efficient acceleration  
than the one considered by OS93. 
Finaly, we would like to note a novel approach
to the shock acceleration by  Vainio \& Schlickeiser (1999), who discuss
conditions allowing  for the shock generated flat spectra without action
of the second order process.

\begin{acknowledgements}    
I am grateful to the anonymous referee  and to Horst Fichtner 
for valuable remarks and corrections. 
The present work was supported by the KBN grant PB 179/P03/96/11.    
\end{acknowledgements}    
    
\section*{References}    
{    
\parskip=0pt    
\parindent=1cm    
\noindent    
Chevalier R.A., 1999, ApJ, in press (ASTRO-PH/ \par 9805315) \\    
Dr\"oge W., Lerche I., Schlickeiser R., 1987, A\&A., 178, 252 \\    
Drury L.O'C., 1983, Space Sci.Rev., 36, 57                 \\    
Hartquist T.W., Morfill G.E., 1983, ApJ, 266, 271                \\    
Hartquist T.W., Morfill G.E., 1984, ApJ, 287, 194                \\    
Micha{\l}ek G., Ostrowski M., 1996, Nonlinear Proc. Geo- \par phys., 3, 66  \\   
Micha{\l}ek G., Ostrowski M., Schlickeiser R., 1998, Solar \par Physics, in    
press  \\    
Ostrowski M., Schlickeiser R., 1993, A\&A, 268, 812  ($\equiv$ \par OS93) \\    
Ostrowski M., 1994, A\&A., 283, 344                          \\    
Schlickeiser R., F\"urst E., 1989, A\&A, 219, 192           \\    
Schlickeiser R., Miller J.A., 1998, ApJ, 492, 352           \\    
Skilling J., 1975, MNRAS, 172, 557           \\    
Vainio R., Schlickeiser R., 1999, A\&A, (in press)   \\
}    
    
\end{document}